%% file: main_eusipco_final.tex
\documentclass[conference]{IEEEtran}
\IEEEoverridecommandlockouts

\usepackage{cite}
\usepackage{amsmath,amssymb,amsfonts,pgfplots,hyperref}
\usepackage{algorithmic}
\usepackage{graphicx}
\usepackage{textcomp}
\usepackage{xcolor}
\def\BibTeX{{\rm B\kern-.05em{\sc i\kern-.025em b}\kern-.08em
    T\kern-.1667em\lower.7ex\hbox{E}\kern-.125emX}}

\usepackage{enumitem}
\setlist{leftmargin=10pt}

\input{./Definitions.tex}

\begin{document}

\title{Cross-Predictive Sparse Bayesian Learning with Application to XL-MIMO Channel Estimation}
\author{
\IEEEauthorblockN{Arttu Arjas and Italo~Atzeni}
\IEEEauthorblockA{Centre for Wireless Communications, University of Oulu, Finland \\
E-mail: \{arttu.arjas, italo.atzeni\}@oulu.fi}
\thanks{This work was supported by the Research Council of Finland (336449 Profi6, 348396 HIGH-6G, and 369116 6G~Flagship).}}

\maketitle

\begin{abstract}
Accurate channel estimation is a key requirement in extremely large-scale multiple-input multiple-output (XL-MIMO) systems. Sparse Bayesian learning (SBL) is a well-established framework for exploiting channel sparsity, but its performance depends on parametric prior assumptions and hyperparameter optimization based on marginal likelihood, which may be sensitive to noise, limited pilot observations, and model mismatch. In this work, we propose \textit{cross-predictive SBL (CP-SBL)}, a data-driven variant of SBL in which the sparsity-inducing weights are learned by minimizing a randomized cross-predictive objective rather than through likelihood maximization. The proposed method preserves the hierarchical Bayesian structure of SBL while replacing parametric prior learning with a predictive consistency criterion derived from random data splitting. Numerical results for near-field XL-MIMO channel estimation show that CP-SBL consistently achieves lower normalized mean squared error than the baseline SBL across a wide range of signal-to-noise ratios, pilot lengths, numbers of antennas, and numbers of propagation paths, with comparable complexity and without requiring manual hyperparameter tuning.
\end{abstract}

\section{Introduction} \label{sec:intro}

Accurate channel estimation is critical in massive multiple-input multiple-output (MIMO) systems, where large-scale antenna arrays are deployed to enable enhanced spatial multiplexing and beamforming \cite{Bjo24}. The ongoing evolution towards extremely large-scale MIMO (XL-MIMO) further intensifies these requirements. By leveraging electrically large apertures and operating at high carrier frequencies (e.g., from FR3 to sub-THz bands \cite{Atz25}), XL-MIMO systems employ even greater antenna counts and exhibit reduced coherence times. This combination increases the dimensionality of the channel while limiting the available pilot resources, making conventional channel estimation approaches inefficient and motivating the development of scalable and robust techniques.

Until recently, most channel models for massive MIMO relied on the far-field assumption, under which impinging wavefronts are approximated as planar. This approximation significantly simplifies channel modeling, as the resulting array responses depend only on angular parameters and become independent of the propagation distance. Consequently, far-field channels admit a sparse representation in the angular domain, a property that has been widely exploited for parametric channel estimation and beamforming design \cite{berger2010application}. However, in emerging XL-MIMO systems, this assumption does not generally hold. In the near-field regime, the wavefronts arriving at the array are inherently spherical, introducing coupling between angles and distances \cite{yang2025near}. As a result, channel sparsity manifests not in the angular domain but in the polar (angle-distance) domain, fundamentally altering the channel structure and necessitating new modeling and estimation approaches.

Exploiting the inherent sparsity of wireless propagation for channel estimation naturally aligns with compressive sensing (CS) frameworks, which seek to recover sparse signals from linear measurements \cite{berger2010application}. Popular CS-based estimators include $\ell_1$-regularized regression \cite{schniter2014channel}, greedy methods like orthogonal matching pursuit \cite{lee2016channel}, and Bayesian techniques such as sparse Bayesian learning (SBL) \cite{tipping2001sparse,prasad2014joint,arjas2025enhanced,Arj26}. Among these, SBL is a robust choice due to its sparsity regulation via hierarchical priors. However, a key limitation of SBL lies in its reliance on priors that are chosen primarily for analytical tractability rather than fidelity to the underlying data distribution \cite{cheng2022rethinking,tipping2001sparse}. Specifically, the inverse-gamma prior commonly used in SBL assumes a heavy-tailed distribution over coefficient magnitudes, which may not align with the actual channel statistics. In such cases, the mismatch can lead to over-shrinkage or underestimation of meaningful components. Moreover, hyperparameters governing sparsity are often tuned manually, which may lead to suboptimal performance and poor adaptability across different scenarios. These issues motivate the need for SBL variants that can adapt to the measurements in a data-driven way.

To overcome these limitations, we propose a novel training strategy for SBL based on randomized cross-prediction. Instead of optimizing the sparsity-inducing weights solely through marginal likelihood maximization, we introduce a predictive loss that encourages better generalization. Related cross-validation and predictive model selection strategies for sparse reconstruction have previously been investigated, e.g., in \cite{ward2009compressed}. A recent work has also explored Stein’s unbiased risk estimator formulations in the context of SBL \cite{slock2022sparse}. In our approach, the observations are randomly partitioned into two disjoint subsets at each iteration. A channel estimate is first obtained from one subset using the current weights, and its predictive performance is then evaluated by assessing how well it explains the measurements in the other subset. The resulting predictive error is used to update the weights. This objective acts as a regularizer on the hyperparameter learning process, guiding the model toward priors that are better aligned with the ultimate goal of accurate channel estimation. The resulting method, which we call \textit{cross-predictive SBL (CP-SBL)}, retains the hierarchical Bayesian structure of the baseline SBL. Unlike prior works that learn priors across multiple data instances \cite{peter2019learned,balaji2025structured}, our approach operates on a single instance and dynamically adjusts the weights based on internal predictive consistency. As a result, it improves estimation accuracy across a wide range of signal-to-noise ratios (SNRs), pilot lengths, and channel sparsity settings, while maintaining computational complexity comparable to that of the baseline SBL. Furthermore, while the performance of the baseline SBL can be highly sensitive to manually selected hyperparameters, the proposed framework operates without any scenario-specific tuning. This removal of empirical calibration enhances robustness, practicality, and reproducibility in real-world deployments. The contributions of this paper are summarized as follows:
\begin{itemize}
    \item We propose CP-SBL, a novel extension of SBL that replaces parameter optimization based on marginal likelihood with a predictive criterion derived from randomized data splitting.
    \item We demonstrate the applicability of CP-SBL to near-field channel estimation, where spherical wavefronts and distance-dependent array responses invalidate conventional far-field sparsity assumptions.
    \item We show that CP-SBL consistently improves channel estimation performance compared with baselines across diverse system configurations.
\end{itemize}

\textit{\textbf{Notations:}} Boldface uppercase and lowercase letters represent matrices and vectors, respectively. $M_{jj}$ denotes the $j$th diagonal entry of matrix $\M$ and $v_j$ the $j$th entry of vector $\v$. For an index set $\setI$ and a matrix $\M \in \Compl^{m \times n}$, $\M(\setI,:) \in \Compl^{|\setI| \times n}$ denotes the submatrix of $\M$ obtained by retaining the rows indexed by $\setI$; likewise, for a vector $\v \in \Compl^{m}$, $\v(\setI) \in \Compl^{|\setI|}$ denotes the subvector of $\v$ formed by retaining the entries indexed by $\setI$.

\section{System Model and Problem Formulation} \label{sec:systemmodel}

We consider uplink channel estimation in an XL-MIMO system, where a base station (BS) equipped with $M$ antennas simultaneously serves $K$ single-antenna user equipments (UEs). The channel vectors between the BS and the UEs are horizontally concatenated to form the channel matrix $\H = [\h_1, \dots, \h_K] \in \Compl^{M \times K}$. To estimate the channel, each UE transmits $N$ predetermined pilot symbols to the BS, collected in the pilot matrix $\P \in \Compl^{K \times N}$. The received signal at the BS is given by
\begin{align} \label{eq:Y}
    \Y = \sqrt{\rho}\H \P + \E \in \Compl^{M \times N},
\end{align}
where $\rho > 0$ is the transmit power and $\E \in \Compl^{M \times N}$ is a matrix of additive white Gaussian noise with i.i.d. $\setC \setN (0,\sigma^2)$ entries.

Assuming near-field propagation, we leverage sparsity in the polar domain. In this setting, the channel matrix can be approximated as
\begin{align}
    \H \approx \F \U,
\end{align}
where $\F \in \Compl^{M \times Q}$ denotes the polar-domain transform matrix \cite{cui2022channel} and $\U \in \Compl^{Q \times K}$ is the corresponding polar-domain channel matrix. Note that the approximation arises because the actual propagation paths may not align exactly with the angle-distance grid defined by $\F$. Finally, vectorizing \eqref{eq:Y} and combining the effects of $\F$ and $\P$ yields
\begin{align}
    \mathrm{vec}(\Y) 
    \approx \mathrm{vec}(\sqrt{\rho} \F \U \P + \E) 
    = \sqrt{\rho}  \A \u + \e \in \Compl^{MN},
\end{align}
with $\A = \P^\tran \otimes \F \in \Compl^{MN \times Q}$, $\u = \mathrm{vec}(\U) \in \Compl^{Q}$, and $\e = \mathrm{vec}(\E) \in \Compl^{MN}$.

Our objective is to estimate the sparse polar-domain channel vector $\u$ from the noisy observation $\Y$ by learning sparsity-inducing weights that promote accurate reconstruction. While SBL leverages the polar-domain sparsity by learning entry-specific weights, its weight prior is typically chosen for mathematical convenience and may not reflect the empirical structure of the data. Against this backdrop, we propose a data-driven weight-learning method that eliminates the need for a parametric prior, producing an effective ``learned prior'' adapted to the observations. Although this work focuses on near-field channel estimation using a polar-domain transform matrix, the proposed framework is general and applicable to arbitrary transform matrices.

\section{Sparse Bayesian Learning}

In this section, we first introduce the baseline SBL with parametric inverse-gamma weight priors. Then, we present the proposed CP-SBL, detailing parameter estimation, key differences relative to the baseline approach, and computational complexity.

\subsection{Baseline SBL} \label{sec:baseline}

SBL places Gaussian priors on the entries of $\u$ with entry-specific weights \cite{tipping2001sparse}. As a baseline, we consider the enhanced version of SBL proposed in \cite{arjas2025enhanced}, termed E-SBL, which places additional scales over the weights to improve the model's flexibility. Denoting the weights by $\w \in \Real_+^Q$ and scales by $\s \in \Real_+^Q$, the prior is expressed as
\begin{align}
    u_j|w_j, s_j \sim \mathcal{N}(0, w_js_j), \quad \forall j = 1, \dots, Q.
\end{align}
Inverse-gamma priors are imposed on the weights and scales, i.e.,
\begin{align}
    \begin{split}
    w_j & \sim \mathcal{IG}\big(\tfrac{\nu}{2}, \tfrac{\nu}{2}\big), \quad \forall j = 1, \dots, Q, \\
    s_j & \sim \mathcal{IG}(\theta, \phi), \quad \forall j = 1, \dots, Q,
    \end{split}
\end{align}
where $\nu, \theta, \phi > 0$ respectively denote the number of degrees of freedom, shape, and scale of the inverse-gamma distribution. The weights and scales are estimated by marginalizing $\u$ out of the model and maximizing the marginal likelihood. This maximization can be performed using the expectation-maximization (EM) algorithm, which iteratively increases the marginal likelihood. At iteration $t$, denoting \mbox{$\R^t = \Diag(\w^t \odot \s^t) \in \Real^{Q \times Q}$}, the EM updates are given by
\begin{align}
    \begin{split}
    w_j^{t+1} &= \frac{\nu/2 + \big(|\mu_j^t|^2 + \Sigma^t_{jj}\big)/s_j^t}{\nu/2 + 2}, \quad \forall j = 1, \dots, Q,\\
    s_j^{t+1} &= \frac{\phi + \big(|\mu_j^t|^2 + \Sigma_{jj}^t\big)/w_j^{t+1}}{\theta + 2}, \quad \forall j = 1, \dots, Q,
    \end{split}
\end{align}
with
\begin{align}
    \begin{split}
    \Sigmab^t &= \left( \frac{1}{\sigma^2} \A^\herm \A + \R^t \right)^{-1} \in \Compl^{Q \times Q}, \\
    \mub^t &= \frac{1}{\sigma^2} \Sigmab^t \A^\herm \y \in \Compl^{Q}.
    \end{split}
\end{align}

\subsection{Proposed Cross-Predictive SBL}

In the baseline SBL, the inverse-gamma weight priors are chosen to promote sparse signal recovery while also providing mathematical convenience, as they yield closed-form updates. When the data significantly deviate from the underlying assumptions, this choice can lead to highly suboptimal performance. To address this, we propose a variant of SBL that learns the weights through a cross-predictive objective, giving rise to an adaptive, nonparametric ``prior'' that better matches the data at hand. The proposed algorithm is an iterative, stochastic procedure for estimating the weights. At each iteration, two subsets of the observed data are generated via randomized splitting. For the first subset, the posterior distribution of $\u$ is reconstructed as conditional Gaussian given the current weights. A cross-predictive objective is then computed, quantifying how well the reconstruction from one subset explains the measurements of the other subset. This objective is then backpropagated to obtain the gradient with respect to the weights, which are subsequently updated via gradient descent.

More specifically, at iteration $t$, we randomly partition the measurement index set $\{1,\dots,MN\}$ uniformly into two disjoint subsets $\setI^t_1$ and $\setI^t_2$ of sizes $B_1 = |\setI^t_1| = \lceil MN/2\rceil$ and $B_2 = |\setI^t_2| = \lfloor MN/2\rfloor$, respectively. Then, we set 
\begin{align}
\y_1^t = \y(\setI_1^t) \in \Compl^{B_1}, \quad & \A_1^t = \A(\setI_1^t,:) \in \Compl^{B_1 \times Q}, \\
\y_2^t = \y(\setI_2^t) \in \Compl^{B_2}, \quad & \A_2^t = \A(\setI_2^t,:) \in \Compl^{B_2 \times Q}.
\end{align}
Since the algorithm does not naturally constrain the weights to be positive, we define the actual weight vector $\widetilde\w \in \Real_+^Q$ in terms of an unconstrained weight vector $\r \in \Real^Q$ with $\widetilde w_j = e^{r_j}, \ \forall j = 1, \dots, Q$, and use $\r$ as the optimization variable. For the first subset, the posterior distribution of $\u$ is given by
\begin{align}
    \u|\y_1^t, \widetilde \w \sim \mathcal{N}(\mub_1^t, \Sigmab_1^t),
\end{align}
with
\begin{align}
    \begin{split}
    \Sigmab_1^t &= \left( \frac{1}{\sigma^2} (\A_1^t)^\herm \A_1^t + \widetilde \W \right)^{-1} \in \Compl^{Q \times Q}, \\
    \mub_1^t &= \frac{1}{\sigma^2} \Sigmab^t (\A_1^t)^\herm \y_1^t \in \Compl^{Q},
    \end{split}
\end{align}
and $\widetilde \W = \Diag(\widetilde \w) \in \Real^{Q \times Q}$. Ideally, the posterior obtained from the first subset should also be consistent with the measurements from the second subset. Enforcing this consistency is conceptually similar to cross-validation, which is commonly used to choose hyperparameters for statistical models by optimizing the predictive performance on unseen data. Motivated by this perspective, we estimate $\r$ by minimizing the cross-predictive objective defined as

\pagebreak

$ $ \vspace{-8mm}

\begin{align}
    \begin{split}
    \textrm{CP}(\r) &= \mathbb{E}_{\u|\y_1^t, \w}\big[\|\A_2^t \u - \y_2^t\|_{2}^2\big] \\
    &= \|\A_2^t \mub_1^t - \y_2^t\|_{2}^2 + \tr\big(\A_2^t \Sigmab_1^t (\A_2^t)^\herm\big).
    \end{split}
\end{align}
The gradient of this objective is computed via backpropagation, and the weights are updated via gradient descent as
\begin{align}
    \r^{t+1} = \r^t - \delta^t \g(\r^t),
\end{align}
where $\g(\r^t) \in \Real^Q$ is the gradient at $\r^t$ and $\delta^t > 0$ is the step size. The $j$th entry of the gradient at $\r$ is given by
\begin{align}
    \begin{split}
    g_j(\r) &= -2\Re\bigg[(\y_2^t)^\herm \A_2^t \Sigmab_1^t \frac{\partial \W}{\partial r_j} \Sigmab_1^t (\A_2^t)^\herm (\A_2^t \mub_1^t - \y_2^t)\bigg] \\
    &\phantom{=} \ - \e_j^\tran \Sigmab_1^t (\A_2^t)^\herm \A_2^t \Sigmab_1^t \e_j \widetilde w_j,
    \end{split}
\end{align}
where $\e_j \in \Real^{Q}$ is the $j$th canonical basis vector of $\Real^Q$. After training, the final estimate is computed as
\begin{align} \label{eq:u_hat}
    \widehat{\u} = \frac{1}{\sigma^2}\left(\frac{1}{\sigma^2}\A^\herm \A + \frac{MN}{B_1}\W^t \right)^{-1} \A^\herm \y \in \Compl^Q.
\end{align}
In \eqref{eq:u_hat}, the factor $\tfrac{MN}{B_1}$ compensates for the scale reduction caused by splitting the data during training: in fact, $(\A_1^t)^\herm \A_1^t$ is smaller in magnitude than $\A^\herm \A$ by a factor of $\tfrac{B_1}{MN}$ (in expectation), so the weights are rescaled accordingly.

The proposed method seeks to minimize the expected validation objective $\mathcal{J}(\r) = \mathbb{E}_{\y_1^t,\y_2^t} \big[\mathrm{CP}(\r)\big]$, where the expectation is taken over random half-data splits. Since the split is resampled independently at each iteration, the resulting weight gradient provides a stochastic approximation to $\nabla_{\r}\mathcal{J}(\r)$. The method may therefore be interpreted as stochastic gradient descent on the expected validation objective, for which standard convergence results apply \cite{kingma2015adam}.

\subsection{Difference Between CP-SBL and Baseline SBL}

CP-SBL differs from the baseline SBL primarily in the criterion used to learn the sparsity-inducing weights. The baseline SBL estimates these weights by maximizing the marginal likelihood under fixed parametric priors, typically chosen for mathematical convenience. While this approach yields closed-form updates, it ties the learning process to prior assumptions that may not align well with practical channel statistics, especially in near-field scenarios.

In CP-SBL, the weights are instead learned through a randomized cross-predictive objective that directly evaluates how well partial reconstructions explain unseen measurements. By repeatedly splitting the observations and enforcing predictive consistency across subsets, CP-SBL guides the weight adaptation toward improved estimation performance rather than likelihood optimization. This results in an implicit data-driven regularization effect without requiring an explicit parametric prior on the weights. Furthermore, the approach avoids the need for manually specified hyperparameters, operating in a fully tuning-free manner.

From a computational standpoint, CP-SBL retains the same dominant complexity order per iteration as the baseline SBL, i.e., $\mathcal{O}(Q^3)$ \cite{tipping2001sparse}, since both methods rely on matrix inversions of comparable sizes. In our numerical results, we observe that CP-SBL typically requires slightly more iterations to converge than the baseline SBL, while maintaining similar overall computational cost. The primary algorithmic difference lies in the update mechanism: CP-SBL replaces EM-based hyperparameter updates with gradient-based optimization driven by predictive error. This modification improves robustness across a range of operating conditions while preserving the overall structure of the SBL framework.

\section{Numerical Results}

In this section, we present numerical results from Monte Carlo simulations demonstrating the superior performance of the proposed CP-SBL compared with the E-SBL algorithm in \cite{arjas2025enhanced} (described in Section~\ref{sec:baseline}) and the variational message passing (VMP)-based approach in \cite{pedersen2012application}. We first outline the simulation setup and evaluation methodology, and then report the performance across different operating conditions.

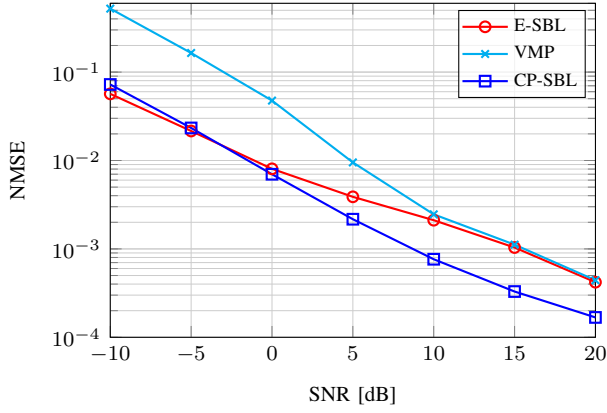
\begin{figure}[!t]
    \centering
    \input{figures/fig_snr}
    \caption{NMSE versus SNR, with $M = 256$, $N = 20$, $K = 5$, and $L = 5$.} \vspace{-2mm}
    \label{fig:nmse_snr}
\end{figure}

\subsection{Simulation Setup}

We evaluate the proposed method in a near-field setting where the BS is equipped with a uniform linear array (ULA) with half-wavelength spacing and the channels follow a clustered multipath model \cite{cui2022channel}, \cite[Ch.~5.6.1]{Bjo24}. Accordingly, the channel of a generic UE (with the UE index omitted for notational simplicity) is generated as
\begin{align}
    \h = \sqrt{\frac{M}{L}}\sum_{l = 1}^L g_l e^{-i\frac{2\pi}{\lambda} r_l} \a(\theta_l, r_l) \in \Compl^M,
\end{align}
where $\lambda$ represents the wavelength, $L$ is the number of propagation paths, and $g_l \in \Compl$, $\theta_l \in \big[ -\frac{\pi}{2}, \frac{\pi}{2} \big]$, and $r_l > 0$ are the small-scale fading coefficient, angle, and distance corresponding to the $l$th path, respectively. Here, $\a(\theta, r) \in \Compl^M$ denotes the near-field ULA steering vector associated with angle $\theta$ and distance $r$ \cite{cui2022channel}. The small-scale fading coefficients $\{ g_l \}_{l=1}^{L}$ are sampled from the complex Gaussian distribution $\mathcal{CN}(0,1)$. Since the ULA steering vector defined as in \cite{cui2022channel} is normalized as $\big\| \a (\theta, r) \big\|_{2}^{2} = M$, we have $\Exp_{g_1, \ldots, g_L}\big[ \| \h \|_{2}^{2} \big] = M$. This normalization allows the (per-antenna) SNR to be defined as $\textrm{SNR} = \tfrac{\rho}{\sigma^2}$. Moreover, the angles $\{ \theta_l \}_{l=1}^{L}$ are sampled from the uniform distribution $\mathcal{U}\big(- \frac{\pi}{4}, \frac{\pi}{4}\big)$, and the propagation distances $\{ r_l \}_{l=1}^{L}$ from the uniform distribution $\mathcal{U}\big(\tfrac{d_{\textrm{F}}}{8},\tfrac{d_{\textrm{F}}}{2}\big)$~m, where $d_{\textrm{F}}=\frac{\lambda}{2} M^2$ denotes the Fraunhofer distance. We fix the Fraunhofer distance and vary the wavelength and number of antennas accordingly. Specifically, we set $d_{\textrm{F}} = 81.92$~m, corresponding to the configurations $M=64$ at carrier frequency of $7.5$~GHz, $M=128$ at $30$~GHz, and $M=256$ at $120$~GHz, thereby spanning carrier frequencies from the FR3 band to the low sub-THz range. Unless otherwise stated, we consider $\textrm{SNR} = 10$~dB, $M=256$ antennas, pilot length $N=20$, $K=5$ users, and $L=5$ propagation paths.

The pilot matrix $\P$ is constructed by selecting the first $K$ rows of the $N$-dimensional DFT matrix. The performance is evaluated using the normalized mean squared error (NMSE), defined as $\textrm{NMSE} = \frac{\mathbb{E}[\|\widehat{\H} - \H\|_\textrm{F}^2]}{\mathbb{E}[\|\H\|_\textrm{F}^2]}$, where $\widehat{\H} \in \Compl^{M \times K}$ denotes the estimated channel matrix. For optimizing the weights in the proposed CP-SBL, we employ Adam \cite{kingma2015adam}, a gradient-descent method that incorporates momentum terms to improve convergence. The step size is fixed to $\delta^t = 0.5$ and the number of iterations to $50$. For E-SBL, we fix the degrees of freedom to $\nu = 1$ and set the shape and scale to $\theta = \phi = 10^{-2}$.

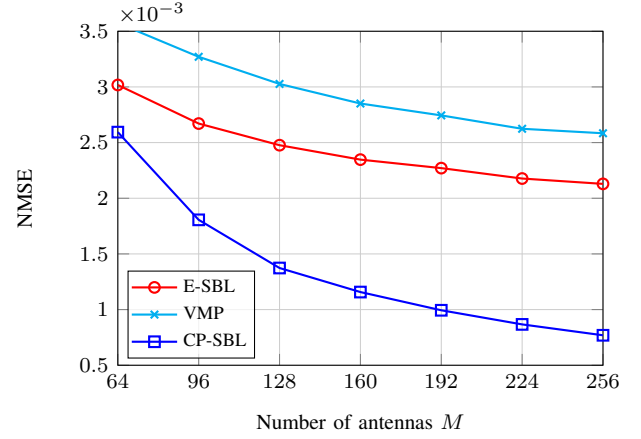
\begin{figure}[!t]
    \centering
    \input{figures/fig_M}
    \caption{NMSE versus $M$, with $\textrm{SNR} = 10$~dB, $N = 20$, $K = 5$, and $L = 5$.} \vspace{-2mm}
    \label{fig:nmse_M}
\end{figure}

\subsection{Results}

Fig.~\ref{fig:nmse_snr} illustrates the NMSE against the SNR. CP-SBL achieves lower NMSE than VMP across the entire SNR range and outperforms E-SBL over most of the SNR range. At low SNR, CP-SBL and E-SBL exhibit comparable performance, while the advantage of CP-SBL becomes clearer as the SNR increases. At $\textrm{SNR} = 10$~dB, CP-SBL attains a $60\%$ reduction in NMSE over E-SBL, and similar trends are observed at higher SNRs. These results indicate that CP-SBL provides improved estimation accuracy over the baselines, particularly in regimes where the noise level is moderate.

Fig.~\ref{fig:nmse_M} shows the NMSE as a function of the number of antennas $M$. As $M$ increases, the NMSE decreases for all methods, reflecting the improved spatial resolution provided by a larger antenna array (i.e., larger polar-domain transform matrix). CP-SBL consistently achieves lower NMSE than the baselines across all considered values of $M$. The performance gap slightly increases with $M$, indicating that CP-SBL more effectively exploits the sparsity structure of near-field channels.

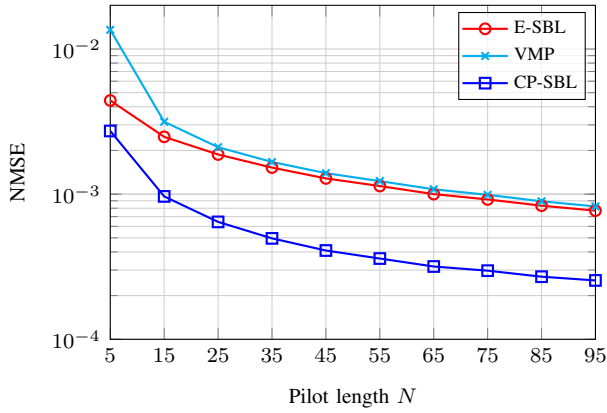
\begin{figure}[!t]
    \centering
    \input{figures/fig_N}
    \caption{NMSE versus $N$, with $\textrm{SNR} = 10$~dB, $M = 256$, $K = 5$, and $L = 5$.} \vspace{-2mm}
    \label{fig:nmse_N}
\end{figure}

Fig.~\ref{fig:nmse_N} shows the NMSE against the pilot length $N$. As $N$ increases, the NMSE of all methods decreases monotonically. Moreover, the relative improvement achieved by CP-SBL over the baselines becomes more pronounced with longer pilots. While the proportional gain is moderate for short pilots, it increases steadily as $N$ grows. This trend indicates that CP-SBL leverages additional pilot resources more effectively, yielding a progressively stronger relative advantage.

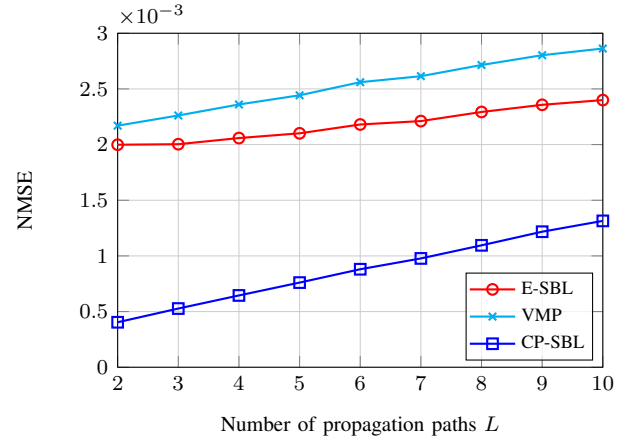
\begin{figure}[!t]
    \centering
    \input{figures/fig_L}
    \caption{NMSE versus $L$, with $\textrm{SNR} = 10$~dB, $M = 256$, $N = 20$, $K = 5$.} \vspace{-2mm}
    \label{fig:nmse_L}
\end{figure}

Lastly, Fig.~\ref{fig:nmse_L} presents the NMSE as a function of the number of propagation paths $L$. As $L$ increases and the channel becomes less sparse, the performance of all methods gradually degrades. For smaller values of $L$, however, CP-SBL achieves a more pronounced performance advantage over the baselines. A similar tendency was observed when increasing the number of antennas, where stronger sparsity conditions also resulted in larger performance gains. Taken together, these results indicate that CP-SBL more effectively captures and leverages the sparsity structure of near-field channels.

\section{Conclusions}

We proposed a fully tuning-free data-driven variant of SBL in which the sparsity-inducing weights are learned by minimizing a randomized cross-predictive objective rather than by marginal likelihood maximization. By replacing parametric prior assumptions with a predictive consistency criterion, the proposed approach adapts the sparsity model to the observed data while preserving the hierarchical Bayesian structure of SBL. Numerical results for near-field XL-MIMO channel estimation showed that our method provides an effective~and practical alternative to the baseline SBL, with comparable~complexity and without requiring manual hyperparameter~tuning.

\bibliographystyle{IEEEtran}
\bibliography{refs_abbr,refs}

\end{document}

%% file: Definitions.tex
\renewcommand{\a}{\mathbf{a}}

\newcommand{\e}{\mathbf{e}}

\newcommand{\g}{\mathbf{g}}
\newcommand{\h}{\mathbf{h}}

\renewcommand{\r}{\mathbf{r}}
\newcommand{\s}{\mathbf{s}}

\renewcommand{\u}{\mathbf{u}}
\renewcommand{\v}{\mathbf{v}}
\newcommand{\w}{\mathbf{w}}

\newcommand{\y}{\mathbf{y}}

\newcommand{\A}{\mathbf{A}}

\newcommand{\E}{\mathbf{E}}
\newcommand{\F}{\mathbf{F}}

\renewcommand{\H}{\mathbf{H}}

\newcommand{\M}{\mathbf{M}}

\renewcommand{\P}{\mathbf{P}}

\newcommand{\R}{\mathbf{R}}

\newcommand{\U}{\mathbf{U}}

\newcommand{\W}{\mathbf{W}}

\newcommand{\Y}{\mathbf{Y}}

\newcommand{\mub}{\boldsymbol{\mu}}

\newcommand{\Sigmab}{\mathbf{\Sigma}}

\newcommand{\setC}{\mathcal{C}}

\newcommand{\setI}{\mathcal{I}}

\newcommand{\setN}{\mathcal{N}}

\newcommand{\Real}{\mbox{$\mathbb{R}$}}
\newcommand{\Compl}{\mbox{$\mathbb{C}$}}

\newcommand{\Diag}{\mathrm{Diag}}

\newcommand{\Exp}{\mathbb{E}}
\newcommand{\herm}{\mathrm{H}}

\renewcommand{\Re}{\mathrm{Re}}

\newcommand{\tr}{\mathrm{tr}}
\newcommand{\tran}{\mathrm{T}}



%% file: figures/fig_snr.tex
\begin{tikzpicture}

\begin{axis}[
	width=8cm,
	height=6cm,
	xmin=-10, xmax=20,
	ymin=1e-4, ymax=0.6,
	xlabel={SNR [dB]},
	ylabel={NMSE},
    scaled y ticks = base 10:-2,
    tick scale binop=\times,
    label style={font=\footnotesize},
    ticklabel style={font=\footnotesize},
	legend style={at={(0.98,0.98)}, anchor=north east},
	legend style={font=\scriptsize, inner sep=1pt, fill opacity=0.75, draw opacity=1, text opacity=1},
	legend cell align=left,
	grid=both,
	grid style={line width=.1pt, draw=gray!40},
    ymode=log,
	title={},
	title style={font=\scriptsize, yshift=-2mm},
]

\addplot[thick, red, mark=o]
table [x=snr, y=SBL, col sep=tab] {figures/txt_files/SNR_final2.txt};
\addlegendentry{E-SBL};

\addplot[thick, cyan, mark=x]
table [x=snr, y=VMP, col sep=tab] {figures/txt_files/SNR_final2.txt};
\addlegendentry{VMP};

\addplot[thick, blue, mark=square]
table [x=snr, y=CP_SBL, col sep=tab]  {figures/txt_files/SNR_final2.txt};
\addlegendentry{CP-SBL};

\end{axis}

\end{tikzpicture}

%% file: figures/fig_M.tex
\begin{tikzpicture}

\begin{axis}[
	width=8cm,
	height=6cm,
	xmin=64, xmax=256,
	ymin=0.0005, ymax=0.0035,
	xlabel={Number of antennas $M$},
	ylabel={NMSE},
	xtick={64,96,...,256},
	ytick={0.0005,0.001,...,0.0035},
    tick scale binop=\times,
    label style={font=\footnotesize},
    ticklabel style={font=\footnotesize},
	legend style={at={(0.02,0.02)}, anchor=south west},
	legend style={font=\scriptsize, inner sep=1pt, fill opacity=0.75, draw opacity=1, text opacity=1},
	legend cell align=left,
	grid=both,
	grid style={line width=.1pt, draw=gray!40},
	title={},
	title style={font=\scriptsize, yshift=-2mm},
]

\addplot[thick, red, mark=o]
table [x=M, y=SBL, col sep=tab] {figures/txt_files/M_final2.txt};
\addlegendentry{E-SBL};

\addplot[thick, cyan, mark=x]
table [x=M, y=VMP, col sep=tab] {figures/txt_files/M_final2.txt};
\addlegendentry{VMP};

\addplot[thick, blue, mark=square]
table [x=M, y=CP_SBL, col sep=tab] {figures/txt_files/M_final2.txt};
\addlegendentry{CP-SBL};

\end{axis}

\end{tikzpicture}

%% file: figures/fig_N.tex
\begin{tikzpicture}

\begin{axis}[
	width=8cm,
	height=6cm,
	xmin=5, xmax=95,
	ymin=0.0001, ymax=0.02,
	xlabel={Pilot length $N$},
	ylabel={NMSE},
	xtick={5,15,25,35,45,55,65,75,85,95},
    scaled y ticks = base 10:-2,
    tick scale binop=\times,
    label style={font=\footnotesize},
    ticklabel style={font=\footnotesize},
	legend style={at={(0.98,0.98)}, anchor=north east},
	legend style={font=\scriptsize, inner sep=1pt, fill opacity=0.75, draw opacity=1, text opacity=1},
	legend cell align=left,
	grid=both,
	grid style={line width=.1pt, draw=gray!40},
    ymode=log,
	title={},
	title style={font=\scriptsize, yshift=-2mm},
]

\addplot[thick, red, mark=o]
table [x=N, y=SBL, col sep=tab] {figures/txt_files/N_final2.txt};
\addlegendentry{E-SBL};

\addplot[thick, cyan, mark=x]
table [x=N, y=VMP, col sep=tab] {figures/txt_files/N_final2.txt};
\addlegendentry{VMP};

\addplot[thick, blue, mark=square]
table [x=N, y=CP_SBL, col sep=tab] {figures/txt_files/N_final2.txt};
\addlegendentry{CP-SBL};

\end{axis}

\end{tikzpicture}

%% file: figures/fig_L.tex
\begin{tikzpicture}

\begin{axis}[
	width=8cm,
	height=6cm,
	xmin=2, xmax=10,
	ymin=0, ymax=0.003,
	xlabel={Number of propagation paths $L$},
	ylabel={NMSE},
	xtick={2,3,...,10},
	ytick={0,0.0005,...,0.003},
    tick scale binop=\times,
    label style={font=\footnotesize},
    ticklabel style={/pgf/number format/fixed,font=\footnotesize},
	legend style={at={(0.98,0.02)}, anchor=south east},
	legend style={font=\scriptsize, inner sep=1pt, fill opacity=0.75, draw opacity=1, text opacity=1},
	legend cell align=left,
	grid=both,
	grid style={line width=.1pt, draw=gray!40},
	title={},
	title style={font=\scriptsize, yshift=-2mm},
]

\addplot[thick, red, mark=o]
table [x=L, y=SBL, col sep=tab] {figures/txt_files/L_final2.txt};
\addlegendentry{E-SBL};

\addplot[thick, cyan, mark=x]
table [x=L, y=VMP, col sep=tab] {figures/txt_files/L_final2.txt};
\addlegendentry{VMP};

\addplot[thick, blue, mark=square]
table [x=L, y=CP_SBL, col sep=tab] {figures/txt_files/L_final2.txt};
\addlegendentry{CP-SBL};

\end{axis}

\end{tikzpicture}